\documentclass[twocolumn]{aastex631}

\usepackage{amssymb} 

\usepackage{amsmath}

\begin{document}

\title{A Closed-Form Analytical Theory of Non-Isobaric Transmission Spectroscopy for Exoplanet Atmospheres}

\correspondingauthor{Leonardos Gkouvelis}

\author[0000-0002-1397-8169]{Leonardos Gkouvelis}
\affiliation{Ludwig Maximilian University, Faculty of Physics, University Observatory,
Scheinerstrasse 1, Munich D-81679, Germany}
\affiliation{Instituto de Astrofísica de Andalucía (IAA-CSIC), 
Glorieta de la Astronomía s/n, E-18008 Granada, Spain}

\begin{abstract}

Analytical models are essential for building physical intuition and guiding the interpretation of exoplanet observations by clarifying the dependencies that shape atmospheric signatures. I present a generalization of the classical isothermal, isobaric transmission model by allowing the opacity to vary with pressure as a power law, $\kappa \propto P^{n}$, and explicitly defining the reference opacity $\kappa_{0}$ at a chosen pressure $P_{0}$. By treating the slant optical depth as an Abel transform of the radial absorption coefficient, I derive a closed-form expression for the effective transit radius in a hydrostatic, isothermal atmosphere with pressure-dependent opacity. The solution provides a compact framework for exploring non-isobaric effects and explicitly links the vertical opacity gradient to observable spectral features. I benchmark the model against empirical transmission spectra of Earth and the hot Jupiter WASP-39b, finding a significantly improved fit relative to the isobaric formula. This generalized expression offers a physically interpretable foundation for analyzing high-precision spectra from JWST and upcoming ARIEL observations, and can serve as a basis for semi-analytical retrieval approaches optimized for computational efficiency.

\end{abstract}


\section{Introduction}\label{sec:intro}

Transmission spectroscopy is one of the key methods for probing exoplanet atmospheres. During the primary transit, a fraction of starlight passes through the planet's limb and allows the tracing of the composition, temperature, and opacity \citep{Brown2001}. This technique has become routine for giant exoplanets, providing detailed insights into their atmospheric processes \citep[e.g.,][]{Espinoza2024,Steinrueck2025}, and is expected to play an important role in characterising Earth-sized rocky planets, particularly with the development of future ultraviolet detectors capable of probing their thermospheres \citep[e.g.,][]{Gkouvelis2025b}. Analytical models give us physical intuition and guide the interpretation of such spectra for more than two decades. The first theoretical framework was established by \citet{Seager2000}, who introduced the concept of an effective transit radius in isothermal, hydrostatic exoplanet atmospheres. \citet{Lecavelier2008} later provided a compact analytical expression linking the transit radius to opacity and scale height, while \citet{Heng2017} refined the formalism and clarified the physical meaning of the effective optical depth criterion for these approximations. Despite these advances, all such models assume an opacity that is independent of pressure. Although this simplification is adopted for tractability \citep{Seager2000, Lecavelier2008, Heng2017}, several studies have emphasized the need to account for vertically varying opacity \citep{Benneke2012, Betremieux2017}.

In reality, many opacity sources, including pressure broadened molecular lines, collision induced absorption, and haze or cloud extinction, exhibit a strong dependence on pressure. This limitation has motivated the inclusion of pressure dependent opacity laws in radiative transfer and emission studies \citep[e.g.,][]{Koll2022}, yet no closed-form analytical solution has been derived for transmission geometry. As a result, the explicit influence of vertical opacity gradients on the effective transit radius has remained concealed within the wavelength-dependent opacity, $\kappa(\lambda)$.  

It is important to note that the non-isobaric effect explored in this work is not the only refinement needed in the analytical transmission models. Non-isothermal temperature profiles or vertically varying chemical abundances can influence transmission spectra and are addressed in numerical radiative-transfer and retrieval frameworks.  Pressure dependent opacity constitutes a structural correction to the effective transit radius formalism itself, and can be incorporated analytically without introducing additional degrees of freedom, in contrast to temperature or compositional gradients which require parametrization. The motivation for focusing on non-isobaric opacity is therefore not that it is uniquely dominant, but that its impact can be quantified independently and treated in closed form within an analytical framework.

In this work, I present a generalized analytical model for the transmission spectrum of an isothermal, hydrostatic atmosphere in which the opacity varies with pressure as a power law, $\kappa \propto P^n$. By treating the slant optical depth as an Abel transform of the radial absorption coefficient, I derive a closed-form expression for the effective transit radius (Section~\ref{sec:derivation}). The resulting formula provides a simple yet flexible framework for exploring non-isobaric effects, capturing physical trends associated with enhanced collisional opacity, compositional gradients, or haze formation. I benchmark the model against empirical transmission spectra of Earth and WASP-39b, finding a clearly improved fit relative to the isobaric formula. This analytical solution extends the foundations laid by \citet{Seager2000}, \citet{Lecavelier2008}, and \citet{Heng2017}, offering a physically interpretable basis for future semi-analytical retrievals and model validation.

\section{Analytical derivation} \label{sec:derivation}

I assume a spherically symmetric, isothermal, hydrostatic atmosphere. The mass density decreases exponentially with radius:
\begin{equation}
\rho(r) = \rho_0 \exp\left( -\frac{r - R_0}{H} \right),
\end{equation}
where $\rho_0$ is the mass density at a reference radius $R_0$, and $H = \frac{k_B T}{\mu m_H g}$ is the pressure scale height, with $g$ the gravitational acceleration,  $T$ the atmospheric temperature, $\mu$ the mean molecular weight, and $m_H$ the mass of the hydrogen atom. Atmospheric opacity generally varies with altitude and pressure, producing vertical gradients that influence the slant optical depth and the effective transit radius. I adopt a power-law dependence of the form
\begin{equation}
\kappa(r) = \kappa_0 \left( \frac{\rho(r)}{\rho_0} \right)^n = \kappa_0 \exp\left( -\frac{n (r - R_0)}{H} \right).
\end{equation}

where the exponent $n$ may vary with wavelength. This formulation captures several physically motivated opacity behaviors observed in planetary atmospheres. Collision-induced absorption (CIA) arises from transient dipoles during molecular collisions and scales with the square of the number density, implying $\kappa_{\mathrm{CIA}} \propto P^2$ for an ideal gas ($n=2$). Pressure-broadened molecular lines exhibit a similar trend, with line-wing opacity typically scaling as $\kappa \propto P^{\beta}$ where $\beta \sim 0.5$–1. In addition, photochemical hazes and condensate clouds can produce opacity that increases with depth if the particle abundance or cross-section grows with pressure, effectively yielding $n>0$ \citep[e.g.,][]{Fortney2005, Morley2015}.   Power-law opacity profiles are also commonly adopted in radiative-transfer and retrieval studies. For instance, \citet{Koll2022} used $\kappa \propto P^n$ to model the infrared photosphere in rocky exoplanet atmospheres. Such parameterizations are both physically motivated and mathematically tractable.  

It follows that  the absorption coefficient is:
\begin{equation}
\alpha(r) = \kappa(r) \cdot \rho(r) = \kappa_0 \rho_0 \exp\left( -\frac{(1+n)(r - R_0)}{H} \right).
\end{equation}
\begin{equation}
\quad \text{I define:}  \quad \alpha(r) = \alpha_0 \exp\left( -\beta (r - R_0) \right),
\end{equation}

\begin{equation}
\quad \text{with} \quad \alpha_0 = \kappa_0 \rho_0, \quad \beta = \frac{1+n}{H}.
\end{equation}
The slant optical depth at impact parameter $r'$ is:
\begin{equation}
\tau(r') = 2 \int_{r'}^\infty \alpha(r) \frac{r}{\sqrt{r^2 - r'^2}} \, \mathrm{d}r.
\end{equation}
This is an Abel transform of $\alpha(r)$.  For $r' \approx R_0$, I substitute $z = r - R_0$:
\begin{equation}   
\tau(r') = 2 \alpha_0 \int_0^\infty e^{-\beta z} \cdot \frac{R_0}{\sqrt{2 R_0 z}} \, \mathrm{d}z 
\label{eq:laplace}
\end{equation}

To  evaluate the integral in equation \ref{eq:laplace} I can make use of the Laplace transform of the inverse square root function which has the  form:
\begin{equation}
\int_0^\infty \frac{e^{-\beta z}}{\sqrt{z}} \, \mathrm{d}z,
\end{equation}
which is a special case of the Laplace transform of the function \( f(z) = z^{-1/2} \):

\begin{equation}
\mathcal{L}\{z^{-1/2}\}(\beta) = \int_0^\infty z^{-1/2} e^{-\beta z} \, \mathrm{d}z.
\end{equation}
This integral is well known and has a closed-form solution  \citet{GradshteynRyzhik2014}\footnote{See also Gradshteyn \& Ryzhik 2014, Entry 3.326.2 for the standard integral.}:
\begin{equation}
\int_0^\infty z^{-1/2} e^{-\beta z} \, \mathrm{d}z = \sqrt{\frac{\pi}{\beta}}.
\end{equation}

so that : 

\begin{equation}
\tau(r') = \alpha_0 \sqrt{2 \pi r'} \cdot \frac{1}{\sqrt{\beta}}.
\end{equation}

Defining $\tau_0 = \tau(R_0)$ and substituting $\alpha_0 = \kappa_0 \rho_0$, I obtain:
\begin{equation}
\tau_0 = \kappa_0 \rho_0 \sqrt{2 \pi R_0} \cdot \sqrt{\frac{H}{1+n}}.
\end{equation}

Using hydrostatic balance, $\rho_0 = \frac{P_0}{g H}$, the expression becomes:
\begin{equation}
\tau_0 = \frac{\kappa_0 P_0}{g H} \cdot \sqrt{2 \pi R_0} \cdot \sqrt{\frac{H}{1+n}} = \frac{\kappa_0 P_0}{g} \cdot \sqrt{ \frac{2\pi R_0}{H (1+n)} }.
\end{equation}

The reference pressure $P_0$ introduced at the reference radius $R_0$ is a mathematical construct used to anchor the asymptotic evaluation of the slant optical depth integral. By construction, it must satisfy $\tau(P_0,\lambda)\gg 1$ at all wavelengths of interest. This reference pressure does not correspond to the pressures directly probed by transmission spectroscopy, which are instead set by the effective optical-depth condition $\tau\sim\mathcal{O}(1)$ (e.g., \citealt{Lecavelier2008}) several scale heights above $P_0$. Once $\tau(P_0)\gg 1$ is satisfied,  the effective transit radius depends only logarithmically on the precise choice of $P_0$ (see \ref{eq:final}). 

\subsection{Effective transit radius}

The transit radius is obtained from the annulus integral:
\begin{equation}
R_p^2 = 2 \int_0^\infty \left[1 - \exp\left(-\tau(r')\right)\right] r' \, \mathrm{d}r'.
\end{equation}

Using the approximation $\tau(r') \approx \tau_0 \exp\left(-\beta (r' - R_0)\right)$  following the same reasoning as \cite{Heng2017} and changing variables $z = r' - R_0$, I obtain :
\begin{equation}
R_p^2 = R_0^2 + 2 R_0 \int_0^\infty \left[ 1 - \exp\left( -\tau_0 e^{-\beta z} \right) \right] \, \mathrm{d}z,
\end{equation}

Changing again variables as: $u = \tau_0 e^{-\beta z} \Rightarrow dz = -\frac{1}{\beta} \frac{du}{u}$ I have:
\begin{equation}
\int_0^\infty \left[1 - \exp\left(-\tau_0 e^{-\beta z}\right)\right] \mathrm{d}z = \frac{1}{\beta} \int_0^{\tau_0} \frac{1 - e^{-u}}{u} \, \mathrm{d}u.
\label{eq:exponantial}
\end{equation}

This integral is a known identity (see  \S28 and \S44 from  \cite{Chandrasekhar1960}; Eq.~3.352 from \citet{GradshteynRyzhik2014}) and has been used in the  literature of the transit theory for exoplanets :

\begin{equation}
\int_0^{\tau_0} \frac{1 - e^{-u}}{u} \, du = \gamma + \ln \tau_0 - E_1(\tau_0),
\end{equation}

where \( \gamma \approx 0.5772 \) is the Euler-Mascheroni constant and \( E_1(\tau_0) = \int_{\tau_0}^\infty \frac{e^{-t}}{t} \, dt \) is the exponential integral of the second kind. In the limit where \( \tau_0 \gtrsim 1 \), the remainder term \( E_1(\tau_0) \) becomes exponentially small:

\begin{equation}
E_1(\tau_0) \sim \frac{e^{-\tau_0}}{\tau_0} \left( 1 - \frac{1}{\tau_0} + \dots \right),
\end{equation}

so that the integral simplifies to the useful approximation:

\begin{equation}
\int_0^{\tau_0} \frac{1 - e^{-u}}{u} \, du \approx \gamma + \ln \tau_0.
\end{equation}

This approximation is accurate for most atmospheric transmission applications, where the slant optical depth \( \tau_0 \) is typically larger than unity. 
Therefore:
\begin{equation}
R_p = R_0 + \frac{H}{1+n} \left( \gamma + \ln \tau_0 \right).
\end{equation}

Substituting $\tau_0$:
\begin{equation}
\boxed{
R_p = R_0 + \frac{H}{1+n} \left[ \gamma + \ln\left( \frac{\kappa_0 P_0}{g} \cdot \sqrt{ \frac{2\pi R_0}{H(1+n)} } \right) \right]
}
\label{eq:final}
\end{equation}
In this final expression, two main differences arise compared to the isobaric formulation. 
First, the factor \( (1+n)^{-1} \) introduces an explicit dependence on the pressure-scaling exponent \(n(\lambda)\),  which modulates the effective atmospheric height and can vary with wavelength depending on the dominant opacity source.  Second, in contrast to the isobaric formulation, the present derivation explicitly defines $\kappa_0$ as the opacity evaluated
at the reference pressure $P_0$ and radius $R_0$. This makes the dependence on local atmospheric conditions explicit, rather than hidden within an arbitrary normalization constant as in the classical isobaric formula. Consequently, the parameters entering Eq.~(21) are physically measurable, $\kappa(\lambda,P_0)$, $P_0$, $g$, and $n(\lambda)$, providing a direct link between laboratory opacity data and observable spectra.

\subsection{Construction of the $n(\lambda)$ function}

In the final expression (Eq.~\ref{eq:final}), the wavelength dependence enters through two quantities: the opacity function \(\kappa_0(\lambda)\), evaluated at the reference pressure \(P_0\), and the pressure dependence exponent \(n(\lambda)\). 
Opacities for individual atmospheric species can be obtained directly from databases such as the Data \& Analysis Center for Exoplanets (DACE)\footnote{\url{https://dace.unige.ch/opacityDatabase/}}.  The DACE database provides wavelength, temperature, and pressure dependent opacities derived from the HITRAN and HITEMP experimental and \textit{ab initio} line lists for each molecular transition \citep{Grimm2015,Grimm2021,Chubb2024}.
For the characterization of the vertical opacity variation I will use the same databases to extract its values.  
I define the wavelength dependent pressure exponent \(n_i(\lambda)\) as the local logarithmic derivative of the opacity with respect to pressure,
\begin{equation}
n_i(\lambda) = \left.\frac{\partial \ln \kappa_i(\lambda,P)}{\partial \ln P}\right|_{T,\lambda}.
\end{equation}
Assuming that, over a limited pressure range, the opacity follows a local power-law dependence \(\kappa_i(\lambda,P) \propto P^{\,n_i(\lambda)}\),  
the above relation integrates to a linear expression in logarithmic form,
\begin{equation}
\ln \kappa_i(\lambda,P) = a_i(\lambda) + n_i(\lambda)\,\ln P ,
\label{eq:linear_n}
\end{equation}
where \(a_i(\lambda)\) is a wavelength-dependent intercept that absorbs the normalization constant. 
This form allows \(n_i(\lambda)\) to be estimated directly from opacity grids by fitting a straight line to \(\ln \kappa_i\) versus \(\ln P\) at fixed temperature.
To assess the validity of the power-law approximation, I evaluated the goodness of fit of $\kappa(\lambda,P)\propto P^{n(\lambda)}$ directly using the opacity database. At fixed wavelength, I performed linear regressions in $\ln\kappa$–$\ln P$ space over the pressure range from the reference pressure $P_0$ down to the lowest pressures the database has to offer. For strong molecular bands, the power-law scaling is nearly exact ($R^2\simeq1$), while in weaker or blended spectral regions the approximation remains accurate at the $>98\%$ level. Deviations from strict linearity reflect transitions between pressure-broadening regimes and indicate that $n(\lambda)$ should be interpreted as an effective exponent capturing the dominant pressure dependence relevant for the transit geometry. This temperature is fixed to the atmospheric  isothermal. 
For each species $i$, I use the opacity grid, which provides $\kappa_i(\lambda, P)$ at discrete pressure levels. At fixed temperature $T$ and for a given resolution $R$, I rebin the opacity spectra to a uniform logarithmic wavelength grid. For each wavelength bin, I fit the linear model from equation \ref{eq:linear_n} where the slope of the fit yields the pressure exponent $n_i(\lambda)$. To reduce noise, the resulting $n_i(\lambda)$ curves are smoothed over a fixed wavelength window (e.g., 0.3 $\mu$m).
The atmosphere is a mixture of species with different abundances and opacities. To compute a representative pressure exponent $n_{\text{mix}}(\lambda)$ for the total opacity field, I compute a weighted average of $n_i(\lambda)$ over species:
\begin{equation}
n_{\text{mix}}(\lambda) = \frac{\sum_i \chi_i \, \kappa_0^i(\lambda, P_0) \, n_i(\lambda)}{\sum_i \chi_i \, \kappa_0^i(\lambda, P_0)} \,,
\end{equation}
where $\chi_i$ is the volume mixing ratio of species $i$ and $\kappa_0^i(\lambda, P_0)$ is its opacity evaluated at a reference pressure $P_0$ (typically 1 bar where $\tau \gg 1$). This formulation gives more weight to species that are both more abundant and more opaque, and reduces to a standard abundance-weighted mean if all opacities are equal.
This effective $n_{\text{mix}}(\lambda)$ is used in the generalized analytic formula for the effective transit radius Equation \ref{eq:final}.

\section{Comparison with Observed Transmission Spectra}

To demonstrate the potential of the analytical formulation, I compare it against empirical transmission spectra of the Earth and the well-studied hot Jupiter WASP-39b. 
Although the Earth's atmosphere is complex, it provides the best possible benchmark because its radius and atmospheric properties are precisely known, allowing us to avoid the normalization degeneracy that typically affects exoplanet transmission observations.

I use the empirical transmission spectrum of Earth from \citet{Macdonald2019}\footnote{The published spectrum is available at DOI: 10.5281/zenodo.8280710 \citep{MacdonaldCowan2019_zenodo}.}, derived from solar occultation measurements by the Atmospheric Chemistry Experiment–Fourier Transform Spectrometer (ACE‐FTS) aboard the Canadian \textit{SCISAT} satellite \citep{Bernath2005, Hughes2014, Boone2019}. 
ACE‐FTS records high‐resolution infrared spectra between 700 and 4400~cm$^{-1}$ (2.27–14.3~$\mu$m) with a spectral resolution of 0.02~cm$^{-1}$ (sampling 0.0025~cm$^{-1}$, $R\!\sim\!10^5$ at 5~$\mu$m). 
\citet{Macdonald2019} integrated these altitude‐resolved transmittance measurements from the surface to the top of the atmosphere, following the approach of \citet{Dalba2015}, to produce an altitude‐integrated and limb‐averaged transmission spectrum analogous to an exoplanet transit. 
They selected cloud‐free sightlines and averaged $\sim$800 spectra across five latitude regions (Arctic, mid‐latitude, and tropical) to obtain a globally averaged, clear‐sky Earth spectrum. 
The resulting dataset has a continuum S/N of $\sim$8000 and probes altitudes down to $\sim$5–10~km, deeper than the refractive limit of a true Earth–Sun analog observation \citep{Misra2014, Betremieux2014}.

To fit the analytical transmission models, I adopted the globally averaged atmospheric composition and temperature structure derived by \citet{Lustig-Yaeger2023}. 
Their retrieval analysis of the empirical Earth spectrum used a mean one-dimensional thermochemical profile consistent with the VPL Earth Model \citep{Robinson2011}, which reproduces the EPOXI Earth flyby observations. 
The adopted composition includes H$_2$O, CO$_2$, O$_3$, O$_2$, N$_2$, CH$_4$, CO, and N$_2$O, along with trace absorbers (HNO$_3$, NO$_2$, CFC-11, and CFC-12), assuming a mean molecular weight of 29~g~mol$^{-1}$, a surface pressure of 1~bar, and a globally averaged temperature of $\sim$255~K. 
These parameters provide a realistic baseline for testing the analytical formulation under clear-sky terrestrial conditions. I adopt $P_0 = 1$~bar not because transmission spectroscopy probes these pressures, but because the continuum slant optical depth at this level is overwhelmingly large across the relevant wavelength range. This choice safely satisfies the asymptotic requirement $\tau(P_0,\lambda) \gg 1$ and ensures numerical stability in the determination of the pressure-dependence exponent $n(\lambda)$. Alternative reference pressures that still fulfill this condition yield indistinguishable effective transit radii within numerical precision. 

\begin{figure*}[ht!]
    \centering
    \includegraphics[scale=0.72, trim=3.3cm 1.5cm 5cm 0cm, clip]{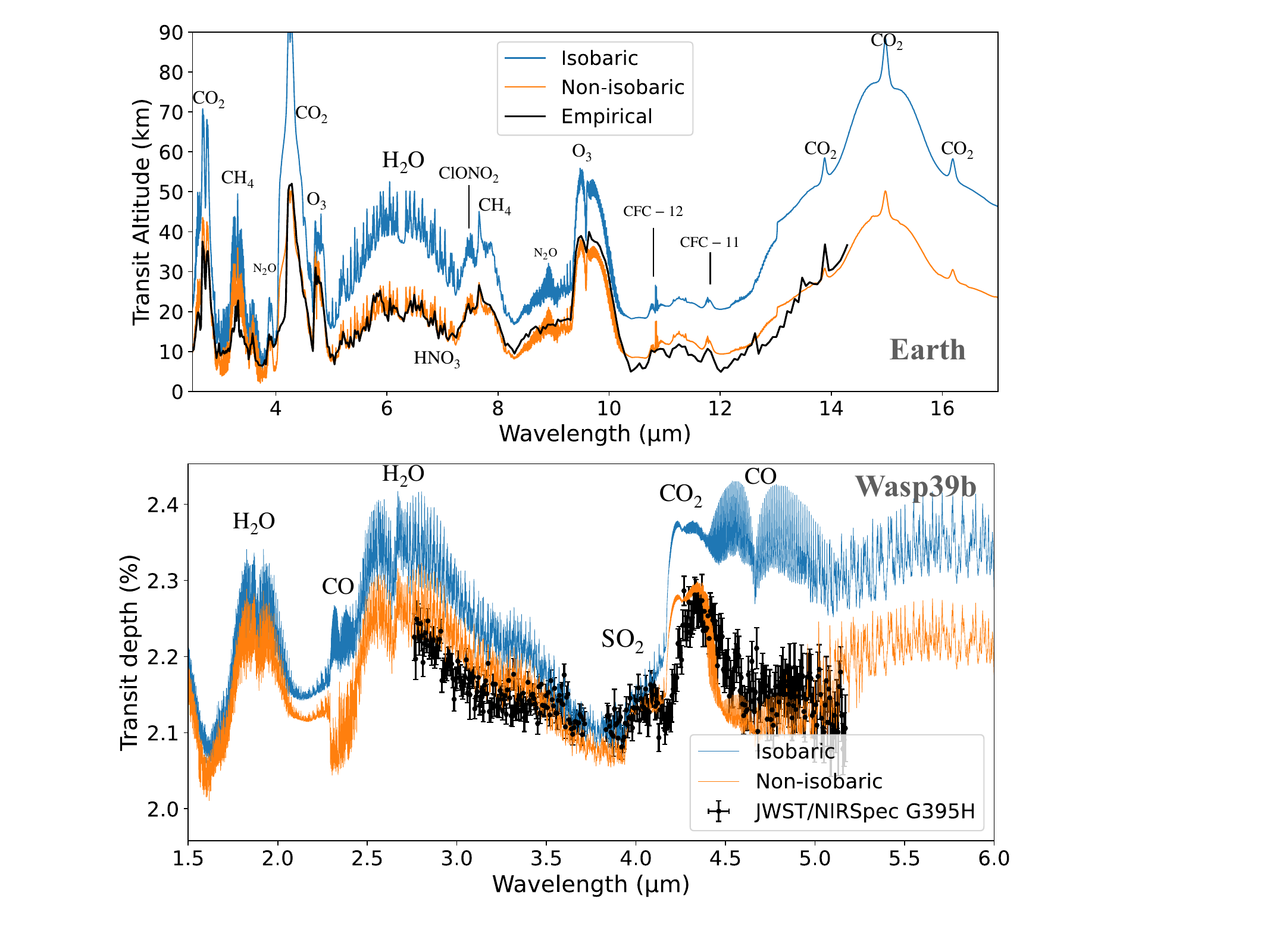}
    \caption{ Comparison between the analytical transmission models and the empirical
spectra. 
\textbf{Top:} Earth's transmission spectrum from Macdonald \&
Cowan~(2019, 2023) derived from ACE-FTS solar-occultation
measurements. The empirical spectrum is shown as a solid black line,
the isobaric analytical model as a blue curve, and the
generalized non-isobaric model as an orange curve. 
\textbf{Bottom:} JWST transmission spectrum of WASP-39b from the ERS
program (Rustamkulov et al.~2023; Tsai et al.~2023), shown as black
points with $1\sigma$ error bars, compared with the same analytical
models (blue: isobaric; orange: non-isobaric). 
In both cases the non-isobaric formulation reproduces the relative
depths and shapes of the major molecular bands, notably the SO$_2$ and
CO$_2$ features, more faithfully than the classical isobaric
approximation.
}
    \label{fig:color}
\end{figure*}

For Earth, there is by construction no degeneracy in the absolute transit radius, as the planetary radius at a given pressure level is known. In particular, the reference radius $R_0$ at $P_0 = 1$~bar is known, allowing the empirical transmission spectrum and the analytical models to be compared directly in absolute altitude space without introducing any free normalization parameters. In the upper panel of Figure~\ref{fig:color}, I overplot the two analytical models, the  isobaric approximation and the generalized non-isobaric formulation, against the empirical transmission spectrum of Earth.  The models are shown at a resolving power of $\approx 500$, while the empirical spectrum is rebinned to $\approx 200$ for clarity.  No vertical normalization or iterative fitting was applied, as my goal is not to achieve the best statistical fit but to illustrate how the non-isobaric model improves upon the isobaric case.  Although the difference between the two models is visually evident, I also performed a statistical analysis to quantitatively assess their relative performance against the empirical data. After rebinning all curves to a common resolving power and evaluating both analytical models using the same fixed atmospheric parameters, the generalized (non-isobaric) formula reduced the root-mean-square deviation from 4.1~km (isobaric model) to 3.0~km and the mean absolute deviation from 3.0~km to 2.3~km.  The corresponding reduced $\chi^2$ decreased from 5.3 (isobaric to 3.7 (non-isobaric), yielding $\Delta{\rm AIC}_c = 608$ in favor of the non-isobaric model indicating an overwhelming improvement.  Here, ${\rm AIC}_c$ denotes the corrected Akaike Information Criterion, which balances model fit and complexity \citep{Burnham2002}.   Although the absolute differences (of order $\sim1$~km) may appear small,  they correspond to a $\sim25\%$ reduction in the typical residual, or $\approx0.15\,H$ for Earth’s atmosphere,  and produce a decrease in the global $\chi^2$ sufficient to yield $\Delta{\rm AIC}_c \approx 600$. 
This indicates that even modest improvements in the altitude domain translate into decisive statistical evidence  when evaluated over hundreds of spectral channels. This constitutes evidence that explicitly accounting for pressure-dependent opacity provides a significantly better match to the empirical spectrum, even without introducing additional free parameters beyond those already present in the classical isobaric formulation.

I further test the analytical solution using the transmission spectrum of the hot Jupiter WASP-39b, one of the best-characterized exoplanets observed with the \textit{James Webb Space Telescope} (JWST) under the Early Release Science (ERS) program.  This target provides an ideal comparison because its atmosphere has been observed across a wide wavelength range with high precision, allowing us to evaluate the applicability of the non-isobaric model under high-temperature, hydrogen-rich conditions that differ markedly from Earth.

I use the combined JWST transmission spectrum reported by the ERS consortium \citep{Rustamkulov2023, Alderson2023, Tsai2023}, which merges data from NIRSpec PRISM, NIRISS SOSS, and NIRCam grism observations. 
Together, these cover 0.5–5.5~$\mu$m at resolving powers ranging from $R\!\sim\!100$ (PRISM) to $R\!\sim\!1500$ (SOSS), yielding a continuous, high signal-to-noise spectrum that clearly reveals multiple molecular bands. 
The observations show prominent absorption from H$_2$O, CO$_2$, and SO$_2$, along with weaker features due to Na and K. 
The detection of SO$_2$ provides the first direct evidence of photochemistry in an exoplanet atmosphere driven by stellar UV irradiation \citep{Tsai2023}. 
I adopt the atmospheric properties retrieved by \citet{Tsai2023}, who derived a hydrogen-dominated atmosphere with mean molecular weight $\mu\!\approx\!2.3$, near-solar to super-solar metallicity, and a terminator temperature of approximately 1100~K. 
The adopted planetary parameters are $R_p\!=\!1.27\,R_{\mathrm{J}}$, $M_p\!=\!0.28\,M_{\mathrm{J}}$, and surface gravity $g\!=\!4.3$~m~s$^{-2}$, providing a consistent physical baseline for testing my analytical model against the JWST spectrum. 
To account for the well-known degeneracy in the absolute transit radius inherent to transmission spectroscopy, I adopt the vertical normalization offset retrieved by \citet{Tsai2023} from their full numerical retrieval analysis. This fixed offset is applied identically to both analytical models, and no additional fitting of the reference altitude is performed.

In the lower panel of Figure~\ref{fig:color}, I compare the isobaric and non-isobaric analytical models to the observed spectrum. 
The non-isobaric solution reproduces the relative depths and shapes of the major absorption bands more faithfully, while the isobaric approximation systematically overestimates the effective transit radius within molecular features. 
In particular, the $\mathrm{SO}_2$ feature is well reproduced by the non-isobaric approach, whereas the isobaric model fails to capture its spectral shape.\\ 
A quantitative comparison yields a reduction in the root-mean-square deviation of the relative transit depth from 0.0358\% to 0.0300\%, and in the mean absolute deviation from 0.0277\% to 0.0229\%. These metrics are computed over all wavelength channels as the RMS and MAD of the residuals between the model and observed wavelength dependent transit depth and expressed as a percentage. The reduced $\chi^2$ decreased from 2.43 to 1.89, yielding $\Delta{\rm AIC}_c = 182$ in favor of the non-isobaric model.  This $\sim$20\% reduction in residuals across 344 spectral channels demonstrates that accounting for pressure dependent opacity improves agreement with the observed spectrum.

\begin{figure*}[ht!]
    \centering
    \includegraphics[scale=0.6, trim=2.3cm 0cm 1cm 0cm, clip]{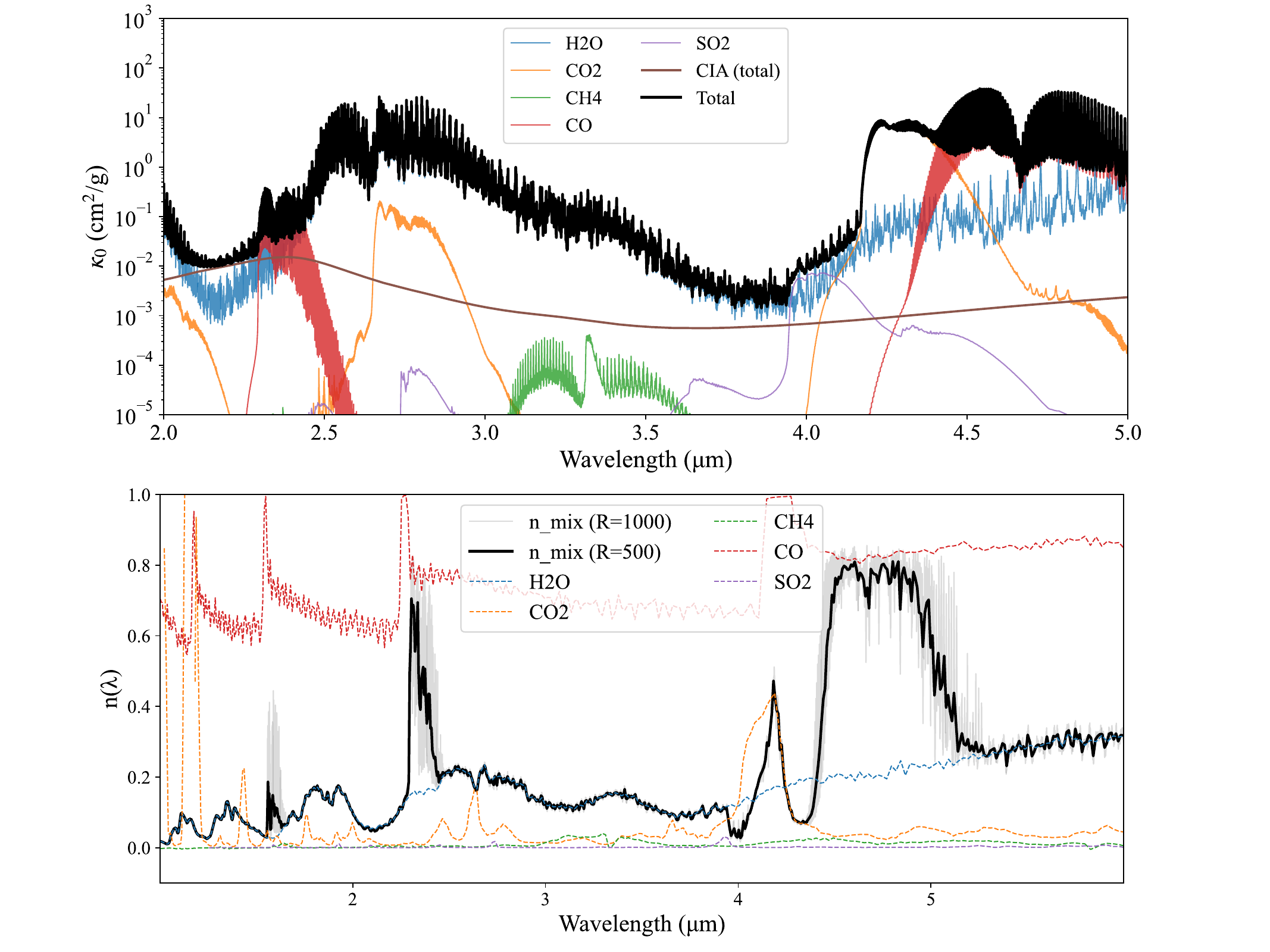}
    \caption{Pressure- and wavelength-dependent opacity behavior for the main
atmospheric absorbers considered in the WASP-39b analysis.
\textbf{Top:} Monochromatic opacities $\kappa_i(\lambda,P_0)$ at
$P_0=1$\,bar and $T=1100$\,K for the dominant species 
(H$_2$O, CO$_2$, CO, SO$_2$, and CH$_4$), computed at
$R=1000$ using the ExoMol/HITEMP/HITRAN line lists.
\textbf{Bottom:} Corresponding pressure-dependence exponents
$n_i(\lambda)=\partial\ln\kappa_i/\partial\ln P|_{T,\lambda}$ and the
abundance-weighted mixture value $n_{\mathrm{mix}}(\lambda)$ for
WASP-39b, smoothed over 0.3\,$\mu$m. 
The results illustrate that $n(\lambda)$ varies strongly with
wavelength and species, reflecting the dominance of different opacity
sources, while the weighted $n_{\mathrm{mix}}(\lambda)$ provides an
effective parameter entering the generalized analytical transmission
formula (Eq.~21).
}
    \label{fig:Wasp39b}
\end{figure*}

To place the fitted exponents in physical context, I summarize the inferred $n(\lambda)$ behavior for both planets compared here. For the terrestrial benchmark, the derived pressure exponent has a median value $\tilde n = 0.75$  over the analyzed wavelength range. The exponent increases in spectral regions that probe deeper atmospheric layers: in the mid-infrared (5-14.3~$\mu$m), which is sensitive to the troposphere, I find median values $\tilde n \approx 0.85-0.88$, consistent with the growing importance of pressure broadened line wings and collision-induced opacity in cool, dense conditions. In contrast, for the hot Jupiter WASP-39b I obtain a substantially smaller median value $\tilde n = 0.12$, with larger values in the 2.5-5.5~$\mu$m range compared to the near-infrared. This difference reflects the higher temperatures, hydrogen-dominated composition, and the different pressure levels sampled in transmission, which enhance the relative contribution of Doppler broadening and line-core opacity. The Earth, WASP-39b comparison therefore provides a useful physical check and supports the interpretation of $n(\lambda)$ as an effective measure of pressure-dependent opacity. Values of $n(\lambda)$ for characteristic spectral bands are shown in table \ref{tab:n_summary}.

\begin{table}[ht]
\centering
\caption{Effective exponent $n(\lambda)$
derived for the Earth and WASP-39b. Reported values are medians,
with the 16-84\% percentile range in parentheses.}
\label{tab:n_summary}
\begin{tabular}{lcc}
\hline\hline
 & Spectral band [$\mu$m] & $\tilde n$  \\
\hline
Earth     & Global        & 0.75 (0.35-0.94) \\
Earth     & 2.3-5.0       & 0.66 (0.30-0.91) \\
Earth     & 5.0-14.3      & 0.85 (0.57-0.96) \\
Earth     & 10.0-14.3     & 0.88 (0.63-0.98) \\
\hline
WASP-39b & Global        & 0.12 (0.06-0.29) \\
WASP-39b & 0.5-2.5       & 0.08 (0.04-0.16) \\
WASP-39b & 2.5-5.5       & 0.17 (0.10-0.48) \\
\hline
\end{tabular}
\end{table}

\section{Conclusions and Discussion} \label{sec:Discussion}

A key insight in deriving analytical expressions for transmission spectra is that the slant optical depth can be written as an {\em Abel transform} of the radial absorption coefficient. The Abel transform is a well‑established mathematical tool used in planetary remote sensing. By expressing the slant optical depth in this form, one benefits from the known analytic and inversion properties of the transform, enabling closed‑form solutions under suitable assumptions (e.g., \citet{Hubert2022}).
This generalized non-isobaric formulation continues the legacy of analytical approaches in exoplanet atmosphere theory. Building upon the foundations laid by \citet{Seager2000} and \citet{Lecavelier2008}, my work adds a flexible and physically motivated generalization that remains analytically tractable. Such formulations not only yield insight into how observables depend on atmospheric structure and composition, but also provide efficient benchmarks for validating numerical radiative transfer models and guiding atmospheric retrievals.
The assumption of a power-law pressure dependence for the opacity, $\kappa \propto P^n$, is physically motivated by the dominant radiative processes in planetary atmospheres. For instance, pressure broadening of molecular lines leads to $\kappa \propto P$, while collision-induced absorption (CIA) scales as $\kappa \propto P^2$ due to its two-body nature. Haze and cloud particle opacities, on the other hand, often decrease with pressure at high altitudes, approximately following inverse power laws. This formulation offers both analytical tractability and physical realism over a broad range of atmospheric conditions. While more complex dependencies are possible, they typically preclude closed-form solutions and offer limited additional insight given current observational constraints.

The generalized expression reveals that the atmospheric height contributing to the transit radius scales inversely with $(1+n)$. This simple factor encapsulates the influence of pressure-dependent opacity, such as that arising from pressure-broadened lines, collision-induced absorption, or vertically varying haze distributions, on the apparent planetary radius. In the limit $n \to 0$, the classical isobaric model is recovered, while positive $n$ values compress the effective atmospheric annulus, mimicking muted spectral features often observed in exoplanet spectra. Comparisons with empirical spectra of Earth and WASP-39b demonstrate that explicitly accounting for non-isobaric opacity significantly improves the match to observations, even without introducing additional free parameters. In both benchmark cases, the analytical models are evaluated under fixed normalization conditions derived either from known planetary properties (Earth) or from independent numerical retrievals (WASP-39b), ensuring that the comparison isolates the effect of pressure-dependent opacity rather than differences in reference altitude. The reduction in residuals and information criteria ($\Delta {\rm AIC}_c \approx 600$ for Earth and $\approx 180$ for WASP-39b) indicates that a simple opacity-pressure scaling captures a substantial portion of the vertical structure encoded in transmission data. 

The formalism naturally extends to include wavelength-dependent exponents $n(\lambda)$ derived from opacity databases, while the reference opacity $\kappa_{0}$ is explicitly tied to the chosen pressure level $P_{0}$, ensuring that both parameters entering the analytical solution are anchored to physically defined opacity values rather than abstract fitting constants. Owing to its compact analytical form and physical transparency, the model provides an efficient foundation for fast forward models and the next generation of semi-analytical and hybrid retrieval techniques.

\begin{acknowledgements}

\textit{Acknowledgments:} I dedicate this work to my former, teacher, Stefanos Trachanas, for striking the spark that lit the fire. The author acknowledges financial support from the Severo Ochoa grant CEX2021-001131-S funded by MCIN/AEI/10.13039/501100011033 and 
Ministerio de Ciencia e Innovación through the project PID2022-137241NB-C43.

\end{acknowledgements}

\appendix

\section{Mathematical limitations}
\label{app:limitations}

\subsection{Limiting case \(n \to -1\)}
The prefactor \(H/(1+n)\) in Eq. \ref{eq:final} diverges as \(n \to -1\), corresponding to a vertical opacity profile \(\kappa \propto P^{-1}\) and therefore to a constant absorption coefficient \(\alpha(r)\) with altitude. 
In this limit, the exponential term in Eq.~(3) flattens and the slant optical depth integral no longer converges, because the atmosphere remains optically thick at all altitudes. 
Physically, this regime represents an atmosphere with no well-defined upper boundary, where starlight would continue to be absorbed indefinitely along the limb. 
Hence, \(n=-1\) marks the boundary between physically admissible opacity scalings (\(n > -1\))—for which the optical depth decreases with height—and non-physical cases (\(n \le -1\)) where the optical depth diverges and the concept of a finite effective transit radius ceases to be meaningful.

\section*{Appendix A.2: Analytical error control}

The analytical expression for the effective transit radius neglects the exponential integral term 
$E_1(\tau_0)$ appearing in the exact formulation (Eq.~\ref{eq:final}). 
For $x > 0$, the exponential integral is defined as
\begin{equation}
E_1(x) = \int_x^{\infty} \frac{e^{-t}}{t}\,dt .
\end{equation}
It satisfies the inequality (\citet{Abramowitz1965}; NIST DLMF~2024\footnote{https://dlmf.nist.gov/})
\begin{equation}
\frac{e^{-x}}{x+1} \leq E_1(x) \leq \frac{e^{-x}}{x}.
\end{equation}
Hence, the neglected contribution to the transit height is bounded by
\begin{equation}
\Delta R = \frac{H}{1+n}\,E_1(\tau_0)
           \leq \frac{H}{1+n}\,\frac{e^{-\tau_0}}{\tau_0}.
\end{equation}
This inequality provides a simple quantitative control of the truncation error:
once $\tau_0 \gtrsim 2$, the remainder term $\Delta R$ becomes exponentially small, 
and the analytical approximation remains accurate to better than a few percent 
of a scale height.

\bibliography{sample631}
\bibliographystyle{aasjournal}

\end{document}